\documentclass[
	conference,
	10pt,
  a4paper
]{IEEEtran}

\IEEEoverridecommandlockouts 
\makeatletter
\def\markboth#1#2{\def\leftmark{\@IEEEcompsoconly{\sffamily}\MakeUppercase{\protect#1}}%
\def\rightmark{\@IEEEcompsoconly{\sffamily}\MakeUppercase{\protect#2}}}
\makeatother

\usepackage[amssymb]{SIunits}
\usepackage[cmex10]{amsmath} 
\usepackage{amssymb, amsthm}
\usepackage{bbm} 
\usepackage{bm}
\usepackage[latin1]{inputenc}
\usepackage{xcolor}
\usepackage{cite}
\usepackage[pdftex]{graphicx}
\usepackage[caption=false, font=footnotesize]{subfig}
\usepackage{url}
\usepackage{mathtools}

\usepackage{xspace} 

\usepackage{booktabs} 

\usepackage{array}
\newcolumntype{L}[1]{>{\raggedright\let\newline\\\arraybackslash\hspace{0pt}}m{#1}}
\newcolumntype{C}[1]{>{\centering\let\newline\\\arraybackslash\hspace{0pt}}m{#1}}
\newcolumntype{R}[1]{>{\raggedleft\let\newline\\\arraybackslash\hspace{0pt}}m{#1}}

\usepackage[
	shortcuts,
	nonumberlist,	
	acronym, 
	hyperfirst=true,
	section=section, 
	order=letter,
	numberedsection=nolabel 
]{glossaries}

\theoremstyle{definition}

\theoremstyle{plain}

\theoremstyle{remark} 
\newtheorem{remark}{Remark}

\newcommand{\FFTm}{\mat{W}}
\DeclareMathOperator{\diag}{\text{diag}}

\newcommand{\define}{\triangleq}
\newcommand{\D}{\mathrm{d}}

\newcommand{\transpose}{\top}

\newcommand{\vect}[1]{\ensuremath{\boldsymbol{#1}}}
\newcommand{\mat}[1]{\ensuremath{\boldsymbol{#1}}}

\newcommand{\imag}{\jmath}
\newcommand{\nlop}[1]{\ensuremath{\bm{\sigma}_{#1}}}
\newcommand{\StepSize}{\ensuremath{\delta}}
\newcommand{\NumSteps}{\ensuremath{M}}

\newcommand{\Rsymb}{\ensuremath{R_\text{s}}}

\newif\ifshow
\showtrue

\newcommand{\abbr}[1]{{#1}}				

\makeatletter
\let\aclOLD=\acl
\renewcommand{\acl}[1]{%
  \begingroup    
  \let\@@underline=\relax
  \aclOLD{#1}%
  \endgroup
}
\makeatother

\newcommand{\NewA}[3]{
	\newacronym{#1}{#2}{#3}
}

\NewA{af}{AF}{\abbr{a}mplify-and-\abbr{f}orward}
\NewA{apsk}{APSK}{\abbr{a}mplitude \abbr{p}hase-\abbr{s}hift \abbr{k}eying}
\NewA{ask}{ASK}{\abbr{a}mplitude-\abbr{s}hift \abbr{k}eying}
\NewA{ase}{ASE}{\abbr{a}mplified \abbr{s}pontaneous \abbr{e}mission}
\NewA{awgn}{AWGN}{\abbr{a}dditive \abbr{w}hite \abbr{G}aussian \abbr{n}oise}
\NewA{biawgn}{BI-AWGN}{binary-input \abbr{a}dditive \abbr{w}hite \abbr{G}aussian \abbr{n}oise}
\NewA{bep}{BEP}{\abbr{b}it \abbr{e}rror \abbr{p}robability}
\NewA{cep}{CEP}{\abbr{c}odeword \abbr{e}rror \abbr{p}robability}
\NewA{ber}{BER}{\abbr{b}it \abbr{e}rror \abbr{r}ate}
\NewA{qap}{QAP}{\abbr{q}uadratic \abbr{a}assignment \abbr{p}roblem}
\NewA{bicm}{BICM}{\abbr{b}it-\abbr{i}nterleaved \abbr{c}oded \abbr{m}odulation}				
\NewA{cm}{CM}{\abbr{c}oded \abbr{m}odulation}				
\NewA{qpsk}{QPSK}{quadrature phase-shift keying}				
\NewA{mlcm}{MLCM}{multilevel coded modulation}				
\NewA{bpsk}{BPSK}{\abbr{b}inary \abbr{p}hase-\abbr{s}hift \abbr{k}eying}				
\NewA{bsc}{BSC}{\abbr{b}inary \abbr{s}ymmetric \abbr{c}hannel}				
\NewA{brgc}{BRGC}{\abbr{b}inary \abbr{r}eflected \abbr{G}ray \abbr{c}ode}				
\NewA{cf}{CF}{\abbr{c}haracteristic \abbr{f}unction}
\NewA{csit}{CSIT}{\abbr{c}annnel \abbr{s}tate \abbr{i}nformation at the \abbr{transmitter}}
\NewA{csi}{CSI}{\abbr{c}annnel \abbr{s}tate \abbr{i}nformation}
\NewA{df}{DF}{\abbr{d}ecode-and-\abbr{f}orward}
\NewA{fd}{FD}{\abbr{f}ull-\abbr{d}uplex}
\NewA{fft}{FFT}{\abbr{f}ast \abbr{F}ourier \abbr{t}ransform}
\NewA{iid}{IID}{independent and \abbr{i}dentically \abbr{d}istributed}
\NewA{isi}{ISI}{\abbr{i}nter\abbr{s}ymbol \abbr{i}nterference}
\NewA{lb}{LB}{\abbr{l}ower \abbr{b}ound}
\NewA{map}{MAP}{\abbr{m}aximum \abbr{a} \abbr{p}osteriori}
\NewA{mf}{MF}{\abbr{m}odulo-and-\abbr{f}orward}
\NewA{mlan}{MLAN}{\abbr{m}odulo-\abbr{l}attice \abbr{a}dditive \abbr{n}oise}
\NewA{ml}{ML}{\abbr{m}aximum \abbr{l}ikelihood}
\NewA{mmse}{MMSE}{\abbr{m}inimum \abbr{m}ean \abbr{s}quare \abbr{e}rror}
\NewA{nlpc}{NLPC}{\abbr{n}onlinear \abbr{p}hase \abbr{c}ompensation}
\NewA{nlpn}{NLPN}{\abbr{n}onlinear \abbr{p}hase \abbr{n}oise}
\NewA{nc}{NC}{\abbr{n}etwork \abbr{c}oding}
\NewA{ofmd}{OFDM}{\abbr{o}rthogonal \abbr{f}requency-\abbr{d}ivision \abbr{m}ultiplexing}			
\NewA{dp}{DP}{\abbr{d}ual-\abbr{p}olarization}
\NewA{pam}{PAM}{\abbr{p}ulse \abbr{a}mplitude \abbr{m}odulation}
\NewA{pdf}{PDF}{\abbr{p}robability \abbr{d}ensity \abbr{f}unction}
\NewA{plnc}{PNC}{\abbr{p}hysical-layer \abbr{n}etwork \abbr{c}oding}
\NewA{psk}{PSK}{\abbr{p}hase-\abbr{s}hift \abbr{k}eying}
\NewA{pmd}{PMD}{\abbr{p}olarization \abbr{m}ode \abbr{d}ispersion}
\NewA{pm}{PM}{\abbr{p}olarization-\abbr{m}ultiplexed}
\NewA{pdm}{PDM}{\abbr{p}olarization-\abbr{d}ivision \abbr{m}ultiplexing}
\NewA{qam}{QAM}{\abbr{q}uadrature \abbr{a}mplitude \abbr{m}odulation}
\NewA{sqp}{SQP}{\abbr{s}equential \abbr{q}uadratic \abbr{p}rogramming}
\NewA{rd}{RD}{\abbr{r}adii \abbr{d}istribution}
\NewA{sep}{SEP}{\abbr{s}ymbol \abbr{e}rror \abbr{p}robability}
\NewA{ser}{SER}{\abbr{s}ymbol \abbr{e}rror \abbr{r}ate}
\NewA{si}{SI}{\abbr{s}ide \abbr{i}nformation}
\NewA{sp}{SP}{\abbr{s}ingle-\abbr{p}olarization}
\NewA{sanr}{SNR}{\abbr{s}ignal-to-(additive-)\abbr{n}oise \abbr{r}atio}
\NewA{snr}{SNR}{\abbr{s}ignal-to-\abbr{n}oise \abbr{r}atio}
\NewA{snlse}{sNLSE}{\abbr{s}tochastic \abbr{n}onlinear \abbr{S}chr\"odinger \abbr{e}quation}
\NewA{nlse}{NLSE}{\abbr{n}onlinear \abbr{S}chr\"odinger \abbr{e}quation}
\NewA{stwrc}{sTRC}{\abbr{s}eparated \abbr{t}wo-way \abbr{r}elay \abbr{c}hannel}
\NewA{stwtrc}{sTTRC}{\abbr{s}eparated \abbr{t}wo-way \abbr{t}wo-\abbr{r}elay \abbr{c}hannel}
\NewA{ts}{TS}{\abbr{t}wo-\abbr{s}tage}
\NewA{twrc}{TRC}{\abbr{t}wo-way \abbr{r}elay \abbr{c}hannel}
\NewA{twtrc}{TTRC}{\abbr{t}wo-way \abbr{t}wo-\abbr{r}elay \abbr{c}hannel}
\NewA{wdm}{WDM}{\abbr{w}avelength-\abbr{d}ivision \abbr{m}ultiplexing}
\NewA{vn}{VN}{variable node}
\NewA{cn}{CN}{constraint node}
\NewA{de}{DE}{density evolution}
\NewA{sc}{SC}{spatially-coupled}
\NewA{ldpc}{LDPC}{low-density parity-check}
\NewA{bp}{BP}{belief propagation}
\NewA{dm}{DM}{dispersion-managed}
\NewA{spm}{SPM}{self-phase modulation}
\NewA{xpm}{XPM}{cross-phase modulation}
\NewA{fwm}{FWM}{four-wave-mixing}
\NewA{ixpm}{IXPM}{intrachannel cross-phase modulation}
\NewA{ifwm}{IFWM}{intrachannel four-wave-mixing}
\NewA{ssfm}{SSFM}{split-step Fourier method}
\NewA{fec}{FEC}{forward error correction}
\NewA{psd}{PSD}{power spectral density}
\NewA{ook}{OOK}{on-off keying}
\NewA{smf}{SMF}{single-mode fiber}
\NewA{ssmf}{SSMF}{standard single-mode fiber}
\NewA{dbp}{DBP}{digital backpropagation}
\NewA{edfa}{EDFA}{erbium-doped fiber amplifier}
\NewA{ofdm}{OFDM}{orthogonal frequency division multiplexing}
\NewA{exit}{EXIT}{extrinsic information transfer}
\NewA{pexit}{P-EXIT}{protograph extrinsic information transfer}
\NewA{osnr}{OSNR}{optical signal-to-noise ratio}
\NewA{roadm}{ROADM}{reconfigurable optical add-drop multiplexer}
\NewA{rps}{RPS}{Raman pump station}
\NewA{mlse}{MLSE}{maximum likelihood sequence estimation}
\NewA{dcf}{DCF}{dispersion compensating fiber}
\NewA{tcm}{TCM}{trellis coded modulation}
\NewA{edc}{EDC}{electronic dispersion compensation}
\NewA{ldpcc}{LDPCC}{low-density parity-check convolutional}
\NewA{llr}{LLR}{log-likelihood ratio}
\NewA{ra}{RA}{repeat-accumulate}
\NewA{ira}{IRA}{irregular-repeat-accumulate}
\NewA{ara}{ARA}{accumulate-repeat-accumulate}
\NewA{mi}{MI}{mutual information}
\NewA{vdmm}{VDMM}{variable degree matched mapping}
\NewA{gmi}{GMI}{generalized mutual information}
\NewA{wd}{WD}{windowed decoder}
\NewA{gn}{GN}{Gaussian noise}
\NewA{hd}{HD}{hard-decision}
\NewA{hdd}{HDD}{hard-decision decoding}
\NewA{sd}{SD}{soft-decision}
\NewA{sdd}{SDD}{soft-decision decoding}
\NewA{emp}{EMP}{extrinsic message passing}
\NewA{imp}{IMP}{intrinsic message passing}
\NewA{gldpc}{GLDPC}{generalized low-density parity-check}
\NewA{scgldpc}{SC-GLDPC}{spatially-coupled generalized low-density parity-check}
\NewA{scldpc}{SC-LDPC}{spatially-coupled low-density parity-check}
\NewA{scfec}{SC-FEC}{spatially-coupled forward error correction}
\NewA{tbbc}{TBBC}{tightly-braided block code}
\NewA{gpc}{GPC}{generalized product code}
\NewA{otn}{OTN}{optical transport network}
\NewA{bch}{BCH}{Bose--Chaudhuri--Hocquenghem}
\NewA{rs}{RS}{Reed--Solomon}
\NewA{bbbc}{BBBC}{block-wise braided block code}
\NewA{hpc}{HPC}{half-product code}
\NewA{pc}{PC}{product code}
\NewA{rv}{RV}{random variable}
\NewA{mbp}{MBP}{multiple block permutator}
\NewA{cer}{CER}{codeword error probability}
\NewA{oh}{OH}{overhead}
\NewA{bdd}{BDD}{bounded-distance decoding}
\NewA{ncg}{NCG}{net coding gain}
\NewA{gwbp}{Galton--Watson branching process}{Galton--Watson branching process}

\newacronym[%
	longplural={binary erasure channels},%
	shortplural={BECs}%
]{bec}{BEC}{binary erasure channel}%

\begin{document}

\title{Deep Learning of the Nonlinear Schrödinger Equation in
Fiber-Optic Communications}


\author{
	\IEEEauthorblockN{
	Christian Häger\IEEEauthorrefmark{1}\IEEEauthorrefmark{2} and
	Henry D.~Pfister\IEEEauthorrefmark{2}
	\thanks{Author e-mails: christian.haeger@chalmers.se and
	henry.pfister@duke.edu.
	This work is part of a project that has received funding
	from the European Union's Horizon 2020 research and innovation
	programme under the Marie Sk\l{}odowska-Curie grant 
	No.~749798. The work was also supported by the National
	Science Foundation (NSF) under grant No.~1609327. Any opinions,
	findings, recommendations, and conclusions expressed in this
	material are those of the authors and do not necessarily reflect the
	views of these sponsors.}}

	\IEEEauthorblockA{\IEEEauthorrefmark{1}%
	Department of Electrical Engineering,
	Chalmers University of Technology,
	Gothenburg, Sweden }
	\IEEEauthorblockA{\IEEEauthorrefmark{2}%
	Department of Electrical and Computer Engineering, Duke University,
	Durham, North Carolina }
}

\maketitle


\begin{abstract}
An important problem in fiber-optic communications is to invert the
nonlinear Schrödinger equation in real time to reverse the
deterministic effects of the channel. Interestingly, the popular
split-step Fourier method (SSFM) leads to a computation graph that is
reminiscent of a deep neural network. This observation allows one to
leverage tools from machine learning to reduce complexity. In
particular, the main disadvantage of the SSFM is that its complexity
using $M$ steps is at least $M$ times larger than a linear equalizer.
This is because the linear SSFM operator is a dense matrix. In
previous work, truncation methods such as frequency sampling,
wavelets, or least-squares have been used to obtain ``cheaper''
operators that can be implemented using filters. However, a large
number of filter taps are typically required to limit truncation
errors. For example, Ip and Kahn showed that for a 10 Gbaud signal and
2000 km optical link, a truncated SSFM with 25 steps would require
70-tap filters in each step and 100 times more operations than linear
equalization. We find that, by jointly optimizing all filters with
deep learning, the complexity can be reduced significantly for similar
accuracy. Using optimized 5-tap and 3-tap filters in an alternating
fashion, one requires only around 2--6 times the complexity of linear
equalization, depending on the implementation.
\end{abstract}


\glsresetall

\section{Introduction}

\begin{figure}[b]
	\includegraphics{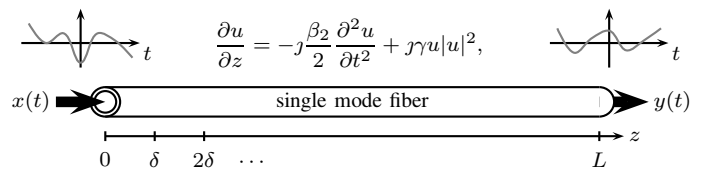}
	\caption{Conceptual signal evolution in a single-mode fiber. The
	nonlinear Schrödinger equation implicitly describes the relationship
	between the input signal $x(t) = u(z=0,t)$ and the output signal
	$y(t) = u(z=L,t)$. The parameters $\beta_2$ and $\gamma$ are,
	respectively, the chromatic dispersion coefficient and the nonlinear
	Kerr parameter of the fiber. The loss term $\alpha u/2$, where
	$\alpha$ is the attenuation parameter, is ignored for simplicity.}
	\label{fig:fiber}
\end{figure}

In a single-mode optical fiber, narrowband signals propagate according
to the nonlinear Schrödinger equation (NLSE)
\cite[p.~40]{Agrawal2006}. This is schematically illustrated in
Fig.~\ref{fig:fiber}. In the absence of noise, the transmitted signal
can thus be recovered by solving an initial value problem (IVP) using
the received signal as a boundary condition. In practice, the received
signal first passes through an analog-to-digital converter and the IVP
can then be solved via receiver digital signal processing (DSP). This
approach is referred to as digital backpropagation (DBP)
 and was inspired by
a similar idea where optical components were used for the processing
\cite{Pare1996}. DBP was first studied as a transmitter pre-distortion
technique \cite{Essiambre2005, Roberts2006}.

%


A major issue with DBP is the large computational burden associated
with a real-time DSP implementation. Thus, various techniques have
been proposed to reduce its complexity \cite{Ip2008, Du2010,
Shen2011,Rafique2011a, Napoli2014, Jarajreh2015,
Secondini2016,Fougstedt2017,Fougstedt2017b}. In essence, the task is
to approximate the solution of a partial differential equation using
as few computational resources as possible. We approach this problem
from a machine-learning perspective. In contrast to, e.g., 
\cite{Monterola2001, Shen2011, Jarajreh2015}, we
focus on deep learning and deep neural networks (NNs), which have
attracted tremendous interest in recent years \cite{LeCun2015}. Our
approach is to obtain a multi-layer computation graph similar to a
deep NN by applying the split-step Fourier method (SSFM)
\cite{Agrawal2006}. This can be seen as an example of a more general
methodology where domain knowledge is used to generate computation
graphs with many layers \cite{Gregor2010}. 


Deep NNs have achieved record-breaking performance for various tasks
such as speech or object recognition \cite{LeCun2015}.  In order to
explain this success, the authors in \cite{Lin2017} argue that most
data of practical interest is generated by some form of hierarchical
or Markov process, often obeying physical principles such as locality
and symmetry. This makes it plausible that there exist efficient
multi-layer computation graphs that can approximate these processes
with few parameters. Our design choices are directly motivated by such
considerations. In particular, our computation graph exploits the
hierarchical problem structure that is introduced by the transmission
process. Moreover, we choose the linear operators in the graph to be
short and symmetric finite impulse response (FIR) filters. 

This paper is a continuation of our work outlined in a recent summary
paper \cite{Haeger2018ofc}. It contains several novel contributions.
Most importantly, we provide a theoretical justification for the deep
learning approach. In particular, while FIR filter design for
chromatic dispersion has been studied extensively in the past
\cite{Savory2008, Ip2008, Zhu2009, Goldfarb2009, Eghbali2014,
Sheikh2016}, the designed filters have shown relatively poor
efficiency when used in a split-step method for DBP due to truncation
errors.  Indeed, we argue that, for computational efficiency, the
filters used in each step should be different and that they should be
optimized \emph{jointly}.  We also compare our approach with multiple
truncation methods (for the filter coefficients) and with ``few-step''
perturbation approaches. 

\section{Digital Backpropagation}

We assume that the signal $x(t)$ is launched into an optical fiber
where it propagates according to the NLSE as shown in
Fig.~\ref{fig:fiber}.  After distance $z=L$, the received signal
$y(t)$ is low-pass (LP) filtered and sampled at $t = k T$ to give a
sequence of samples $\{y_k\}_{k \in \mathbb{Z}}$. Our goal is to
\emph{efficiently} recover the signal $x(t)$ (or a sampled version
thereof) from $\{y_k\}_{k \in \mathbb{Z}}$. 

\subsection{Split-Step Fourier Method}

\begin{figure}[t]
	\centering
		\includegraphics{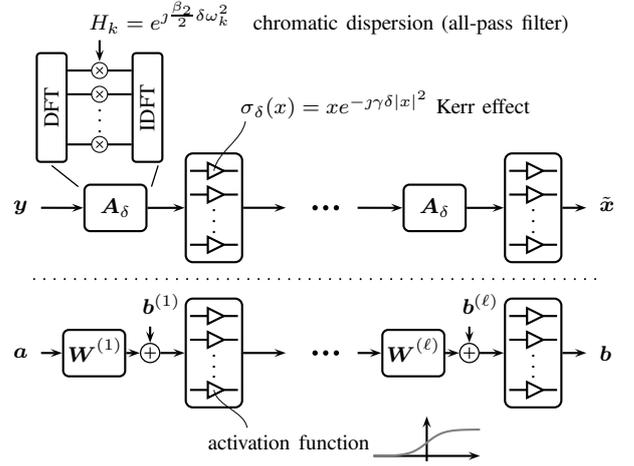}
	\caption{Block diagram of the split-step Fourier method to
	numerically solve the nonlinear Schrödinger equation (top) and the
	canonical model of a deep feed-forward neural network (bottom).}
	\label{fig:SSFM}
\end{figure}

The popular SSFM is based on a block-wise receiver processing. To that
end, assume that we collect $n$ received samples into a vector
$\vect{y} = (y_1, \dots, y_n)^\transpose \in \mathbb{C}^n$. Consider
now the time-discretized NLSE
\begin{align}
	\label{eq:discretized_nlse}
	\frac{\D \vect{u}(z)}{\D z} 
	= \mat{A} \vect{u}(z) +
	\imag \gamma \bm{\rho}(\vect{u}(z)), 
\end{align}
where $\vect{u}(z) \in \mathbb{C}^n$ represents the sampled waveform
at position $z$ along the fiber, $\vect{A} = \FFTm^{-1} \diag(H_1,
\dots, H_n) \FFTm$, $\FFTm$ is the $n\times n$ discrete Fourier
transform (DFT) matrix, $H_k = -\imag \frac{\beta_2}{2} \omega_k^2$,
$\omega_k = 2 \pi f_k$ is the $k$-th DFT angular frequency, and
$\bm{\rho}$ is defined as the element-wise application of $\rho(x) = x
|x|^2$. To derive the SSFM, the fiber is conceptually divided into
$\NumSteps$ segments of length $\StepSize = L / \NumSteps$. Then, it
is assumed that for sufficiently small $\delta$, the effects stemming
from the two terms on the right hand side of
\eqref{eq:discretized_nlse} can be separated.  More precisely, for
$\gamma = 0$, \eqref{eq:discretized_nlse} is linear with solution
$\vect{u}(z) = \mat{A}_z \vect{u}_0$, where $\mat{A}_z \define e^{z
\mat{A}}$. For $\beta_2 = 0$, the solution is $\vect{u}(z) =
\nlop{z}(\vect{u}_0)$, where $\nlop{z}$ is the element-wise
application of $\sigma_z(x) = x e^{\imag \gamma z |x|^2 }$.
Alternating between these two operators for $z=-\delta$ leads to the
block diagram shown in the top part of Fig.~\ref{fig:SSFM}.  

The degree to which the obtained vector ${\tilde{\vect{x}}}$
constitutes a good approximation of $x(t)$ is now a question of
choosing $M$, $T$, and $n$. In practice, $1/T$ is typically an integer
multiple of the baud rate and $n$ is chosen to minimize the overhead
in overlap-and-save techniques for continuous data transmission.
Increasing $M$ leads to a more accurate approximation, but also
increases complexity, as discussed in the next section. 




\subsection{Implementation Complexity and Few-Step Approaches}


Ignoring the complexity of the nonlinear steps, the SSFM can be
implemented using $M$ DFT/IDFT pairs, utilizing the fast Fourier
transform. On the other hand, a linear equalizer can be implemented
using a single DFT/IDFT pair.  Based on this reasoning, the SSFM is at
least $M$ times more complex than linear equalization. This motivates
a number of approaches that focus on reducing the number of steps,
see, e.g., \cite{Du2010, Rafique2011a, Secondini2016} and references
therein. 


\section{Deep Feed-Forward Neural Networks}


Deep feed-forward NNs map an input vector $\vect{a}$ to an output
vector $\vect{b}$ by alternating between affine transformations and
pointwise nonlinearities \cite{LeCun2015}, \cite[Eq.~(6)]{Lin2017}.
This is illustrated in the bottom part of Fig.~\ref{fig:SSFM}. The
matrices $\vect{W}^{(1)}, \dots, \vect{W}^{(\ell)}$ and vectors
$\vect{b}^{(1)},\dots,\vect{b}^{(\ell)}$ are the network weights and
biases, respectively, and $\ell$ is the number of layers. The
nonlinearities typically correspond to some activation function, e.g.,
the logistic or sigmoid function. 

While the similarity between the two computation graphs in
Fig.~\ref{fig:SSFM} is apparent, there are, however, important
differences. The one that is most relevant for this paper is the
sparsity level of the linear operators. In order to be computationally
efficient, deep NNs are typically designed to have very sparse weight
matrices in most of the layers, whereas the linear propagation
operator $\mat{A}_\delta$ in the SSFM is a dense matrix.  

\begin{remark}
	In that regard, one may argue that \eqref{eq:discretized_nlse} is a
	``computationally inefficient'' time-discretization of the NLSE, in
	the sense that it relates local propagation changes to all time
	instances. A different time-discretization approach is via partial
	discretization or finite-difference methods. Indeed,
	finite-difference methods can be more computationally efficient than
	the SSFM in some applications \cite[Sec.~2.4.2]{Agrawal2006}.
	However, to the best of our knowledge, finite-difference methods
	have not been studied for real-time DBP. One reason for this might
	be that many methods that show good performance are implicit, i.e.,
	they require solving a system of equations at each step. This makes
	it challenging to satisfy a real-time constraint. 
\end{remark}
	

\section{Filter Design for Chromatic Dispersion}

\ifshow

According to the NLSE, chromatic dispersion acts as an all-pass filter
with frequency response $H(\omega) = e^{\imag \kappa \omega^2}$, where
$\kappa \define - \beta_2 \delta / (2 T^2)$, $\omega \define 2 \pi f 
T$, and $\delta$ is the transmission distance. Various approaches have
been proposed to approximate this response (over a fixed bandwidth)
with an FIR filter. For example, since the inverse Fourier transform
of $H(\omega)$ can be computed analytically, filter coefficients may
be obtained through direct sampling and truncation \cite{Savory2008}.
Other approaches include frequency-domain sampling (FDS)
\cite{Ip2008}, wavelets \cite{Goldfarb2009}, and least-squares (LS)
\cite{Eghbali2014, Sheikh2016}. 


\subsection{Parameter Efficiency in Split-Step Methods}

Time-domain FIR filtering has been suggested for DBP in, e.g.,
\cite{Ip2008, Zhu2009, Goldfarb2009, Fougstedt2017, Fougstedt2017b}.
In the SSFM, approximating $H(\omega)$ with a short FIR filter can be
interpreted as a truncation of $\mat{A}_\delta$ to obtain a sparse
banded matrix.  

To estimate the required filter length, one may use the fact that
chromatic dispersion leads to a group delay difference of $2 \pi
\beta_2 \Delta f \delta$ over a bandwidth $\Delta f$ and distance
$\delta$. Normalizing by the sampling interval $T$, this confines the
memory to
\begin{align}
	\label{eq:rule_of_thumb}
	K_{\text{cd}} = 2 \pi \beta_2 \Delta f \delta / T
\end{align}
samples. For example, we have $\beta_2 = -21.668$ ps$^2$/km, $\delta =
80\,$km, and $1/T = 32.1$ GHz for the system studied in \cite{Ip2008}.
The receiver bandwidth is $32.1$ GHz, but it is limited by an LP
filter with 3-dB cutoff at $12.4$ GHz. Thus, FIR filters with 4--12
taps should be sufficient. However, 70-tap filters are required to
obtain acceptable accuracy using FDS \cite{Ip2008}. Similar
observation apply to the results in \cite{Goldfarb2009, Fougstedt2017,
Fougstedt2017b, Zhu2009}, i.e., the required filter length is
significantly longer than predicted by \eqref{eq:rule_of_thumb}. 

\subsection{Joint Filter Optimization}
\label{sec:joint_filter_optimization}

\begin{figure}[t]
	\centering
	\subfloat[$\vect{h} \define \vect{h}^{(1)} = \dots
	= \vect{h}^{(M)}$]{\includegraphics{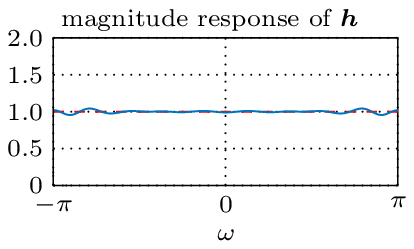}}
	\quad
	\subfloat[$\vect{h} \define \vect{h}^{(1)} * \dots *
	\vect{h}^{(M)}$]{\includegraphics{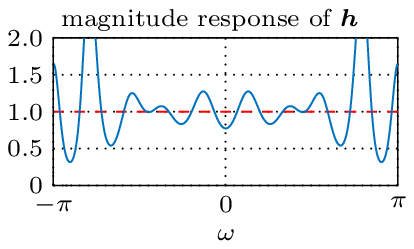}}
	\caption{Schematic illustration of the truncation error when using
	the same (or very similar) FIR filters in a split-step method, where
	$*$ denotes convolution.}
	\label{fig:truncation_error}
\end{figure}


In previous work, a single filter or filter pair is designed and then
used repeatedly in the SSFM. In this case, the truncation error
accumulates coherently, leading to an undesired overall magnitude
response as illustrated in Fig.~\ref{fig:truncation_error}. The effect
is well known and a simple way to control it is by increasing the
filter length. We propose instead to optimize \emph{all} $M$ filters
jointly. 

\begin{remark}
	In \cite{Fougstedt2017, Fougstedt2017b}, filter coefficient
	quantization is studied for time-domain DBP. They highlight the
	effect of correlated quantization errors and propose random
	dithering \cite{Fougstedt2017} and co-optimization of quantization
	levels of filter pairs \cite{Fougstedt2017b}. While this does not
	address the truncation error problem directly, it does alleviate it
	somewhat. 
\end{remark}

In this section, we illustrate how a joint filter optimization can be
done in a way such that the problem admits a (possibly suboptimal)
solution strategy via iteratively solving a set of weighted LS
problems. This approach is simple and provides valuable insight into
the problem. The optimized coefficients are then used as the initial
starting point for the gradient-based deep learning approach discussed
in the next section. 

For simplicity, it is assumed that each of the $M$ filters
$\vect{h}^{(i)} = (h_{-K}^{(i)}, \dots, h_0^{(i)}, \dots,
h_K^{(i)})^\transpose$ for $i = 1, \dots, M$ has $2K+1$ taps. The
generalization to unequal filter lengths is straightforward. Let
$\mathcal{F}(\vect{h}^{(i)}) = \sum_{k=-K}^K h_k^{(i)} e^{- \imag k
\omega}$ be the discrete-time Fourier transform of $\vect{h}^{(i)}$.
We use $\mathcal{F}(\vect{h}^{(i)}) \circeq e^{\imag \kappa \omega^2}$
to denote an objective, i.e., the symbol $\circeq$ may be interpreted
as ``should be close to''. The standard filter design uses the same
objective for each of the $M$ filters, i.e., each filter should
approximate, as closely as possible, the chromatic dispersion transfer
function $e^{\imag \kappa \omega^2}$ over some frequency range. In
this case, one finds that all $M$ filters should be the same. In
particular, after discretizing the problem with $\omega_i = 2 \pi i /
N$ for $i = -N/2, \dots, N/2$, one may use standard techniques to
solve the linear LS problem $\min_{\vect{h}^{(i)}}\| \vect{B}
\vect{h}^{(i)} - \vect{d} \|^2$, where $\vect{d} = (d_{-N/2}, \dots,
d_0, \dots, d_{N/2})^\transpose$ with $d_i \define e^{\imag \kappa
\omega_i^2}$ and $\vect{B}$ is an $(N+1) \times (2K+1)$ DFT matrix. 

On the other hand, by sacrificing some accuracy for the individual
frequency responses, it may be possible to achieve a better combined
response of neighboring filters and also a better overall response.
This leads to the set of objectives
\begin{equation}
	\begin{aligned}
	\label{eq:objectives}
\mathcal{F}(\vect{h}^{(i)}) &\circeq e^{\imag \kappa
\omega^2}, &&i = 1,2, \dots, M\\
\mathcal{F}(\vect{h}^{(i)} * \vect{h}^{(i+1)}) &\circeq e^{\imag 2
\kappa
\omega^2}, &&i = 1,2, \dots, M-1\\
&\,\,\,\vdots && \\
\mathcal{F}(\vect{h}^{(1)} * \cdots * \vect{h}^{(M)}) &\circeq
e^{\imag M \kappa
\omega^2}.
	\end{aligned}
\end{equation}
Keeping the coefficients for all but one filter constant,
\eqref{eq:objectives} can be written as a standard weighted LS problem. Since,
e.g., $\mathcal{F}(\vect{h}^{(i)} * \vect{h}^{(i+1)})
=\mathcal{F}(\vect{h}^{(i)}) \mathcal{F}(\vect{h}^{(i+1)})$, we have
$(\mat{B} \vect{h}^{(i)}) \circ (\mat{B} \vect{h}^{(i+1)})$ in the
discretized problem, where $\circ$ denotes element-wise
multiplication. Hence, one obtains 
\begin{align}
	\label{eq:weighted_ls}
	\min_{\vect{h}^{(i)}} \sum_{j=1}^{O_i} \lambda_j \| (\vect{B}
	\vect{h}^{(i)}) \circ \vect{e}_j - \vect{d}_j \|^2,
\end{align}
where $O_i$ is the number of objectives, $\lambda_j > 0$ are weights,
$\vect{e}_j$ are constant vectors representing the influence of other
filters and $\vect{d}_j$ are the discretized objective vectors.  A
simple strategy for the joint optimization is then to solve
\eqref{eq:weighted_ls} for each of the $M$ filters in an iterative
fashion. The weights $\lambda_1, \dots, \lambda_{O_i}$ can be chosen
based on a suitable system criterion. 

We assume $x(t) = \sum_{k = -\infty}^{\infty} x_k p(t - k/\Rsymb)$,
where $x_k \in \mathbb{C}$ are the data symbols, $p(t)$ is the pulse
shape, and $\Rsymb$ is the baud rate. For the block-wise processing,
the estimated symbol vector $\hat{\vect{x}}$ is obtained by passing
$\vect{z}$ (see Fig.~\ref{fig:SSFM}) through a digital matched filter
(MF) followed by a phase-offset rotation. The mean squared error
$\|\vect{x} - \hat{\vect{x}}\|^2$ is then used as a criterion to be
minimized. Assuming that $\|\vect{x}\|^2$ is constant for all
$\vect{x}$, this is equivalent to maximizing the effective
signal-to-noise ratio (SNR) $\|\vect{x}\|^2 / \|\vect{x} -
\hat{\vect{x}}\|^2$. 


\fi


\section{Learned Digital Backpropagation}
\label{sec:ldbp}

\ifshow

In \cite{Haeger2018ofc}, we have proposed to use deep learning for the
joint filter optimization. The resulting method is referred to as
learned DBP (LDBP). For LDBP, the computation graph of the SSFM is
modified by interpreting all matrices $\mat{A}_\delta$ as tunable
parameters corresponding to the filters $\vect{h}^{(1)}, \dots,
\vect{h}^{(M)}$, similar to the weight matrices in a deep NN. The
nonlinearities are changed to $\smash{\bm{\sigma}^{(i)} : \mathbb{C}^n
\to \mathbb{C}^n}$ which act element-wise using $\sigma^{(i)}(x) = x
e^{-\imag \gamma_i |x|^2}$, where $\gamma_i \in \mathbb{R}$ is a
tunable parameter.  

The computation graph including the MF and phase-offset rotation is
implemented in TensorFlow. All parameters $\theta = \{\vect{h}^{(1)},
\dots, \vect{h}^{(M)}, \gamma_1, \dots, \gamma_M\}$ are optimized by
using many pairs $(\vect{y}, \vect{x})$ of input and desired--output
examples and adjusting the parameters such that the loss $\|\vect{x} -
\hat{\vect{x}}\|^2$ decreases. For this, we use the built-in
\emph{Adam} optimizer with a mini-batch size of 30 and a fixed
learning rate. To find a good starting point for the filter
coefficients, we employ the LS method described in
Sec.~\ref{sec:joint_filter_optimization}. While it is possible to use
random starting points, we observe that a better final solution can be
obtained with pre-optimized coefficients.  


\section{Results and Discussion}


We revisit the parameters in \cite{Ip2008}, using a different LP
filter and transmit signals. Extensions to wavelength division
multiplexing (WDM) systems and higher baud rates are discussed below.
The optical link consists of $25$ spans of $80\,$km fiber and an
amplifier is inserted after each span to compensate for the signal
attenuation. All parameters are summarized in
Fig.~\ref{fig:results}.\footnote{A Butterworth LP filter and QPSK
modulation are assumed in \cite{Ip2008}.} Forward propagation is
simulated with 6 samples/symbol using the SSFM with 50 steps per span
(StPS), i.e., $M = 1250$. 

LDBP uses 1 StPS (i.e., $M = 25$), alternating between 5-tap and 3-tap
filters. The effective SNR after training is shown in
Fig.~\ref{fig:results} by the green line (triangles). As a reference,
we show the performance of linear equalization (red) and DBP with 1
StPS using frequency-domain filtering (blue). The linear equalizer
uses LS-optimal coefficients with constrained out-of-band gain (LS-CO)
\cite{Sheikh2016}. LDBP achieves a peak SNR of 21.9 dB using $13 \cdot
4 + 12 \cdot 2 + 1 = 77$ total taps. After increasing the filter
lengths to 7 and 5 (127 total taps), one obtains essentially the same
peak SNR as frequency-domain filtering. 

\subsection{Comparison to Other Truncation Methods}

The performance of FDS (circles) and LS-CO (squares) is shown in
Fig.~\ref{fig:results} as a comparison. The same filter is used in
each step and the length is chosen such that the peak SNR is around 22
dB.  For this, 15-tap filters are required for FDS (351 total taps)
and 9-tap filters for LS-CO (201 total taps).  This is roughly 5 and 3
times more than required for LDBP.

In \cite{Ip2008}, 70-tap filters based on FDS are required for similar
accuracy. This is likely due to the higher oversampling factor used (3
samples/symbol). While a higher oversampling factor may increase the
maximum SNR achievable via DBP, it can also adversely affect the
performance if truncation errors are taken into account. In general,
it is difficult to predict how truncation errors affect the SNR in a
nonlinear system. 

\subsection{Complexity Compared to Linear Equalization}

We use multiplications as a surrogate for complexity and assume that
the exponential function is implemented with a look-up table, similar
to \cite{Ip2008}. For the nonlinear steps, one needs to square each
sample (2 real multiplications), multiply by $\gamma_i$, and compute
the phase rotation (4 real multiplications). This gives $25 \cdot 7 =
175$ real multiplications per sample. For the linear steps, one has to
account for 13 filters with 5 taps and 12 filters with 3 taps. All
filters have symmetric coefficients and can be implemented using a
folded structure with $h_0$-normalization as shown in
Fig.~\ref{fig:filter}. This gives $39 \cdot 4 = 156$ real
multiplications per sample. In comparison, the fractionally-spaced
linear equalizer in \cite{Ip2008} requires 188 real multiplications
per data symbol operating at $3/2$ samples/symbol. Thus, LDBP requires
3.5 times more multiplications per symbol. For the same oversampling
factor as LDBP, the linear equalizer has $75$ taps
(cf.~\eqref{eq:rule_of_thumb} with $\Delta f = 1.2 \cdot 10.7$ GHz).
This leads to $38 \cdot 4 = 152$ real multiplications with a folded
implementation. Thus, LDBP requires around 2 times more operations.
If the linear equalizer is implemented in the frequency domain, the
number of real multiplication is reduced to $n(4 \log_2 n + 4)/(n-75)
\approx 50$ per sample (see, e.g., \cite[Sec.~4]{Secondini2016}),
which increases the estimated complexity overhead factor to 6. 

\begin{figure}[t]
	\centering
		\includegraphics{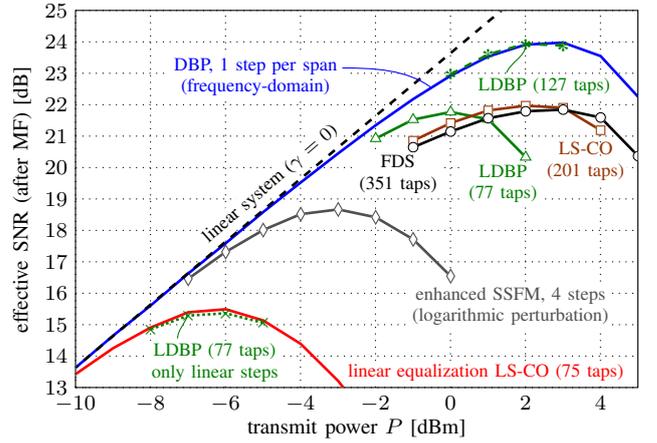}

		\caption{Results for
	$25 \times 80\,$km
		single-mode fiber ($\beta_2 = -21.668\,$ps{\tiny$^2$}/km, $\gamma =
		1.3\,$1/W/km, $\alpha = 0.2\,$dB/km), $5\,$dB amplifier
		noise figure, Gaussian root-raised
		cosine pulses (0.1 roll-off), $\Rsymb = 10.7\,$Gbaud, 2
		samples/symbol ($1/T = 21.4\,$GHz), 15 GHz brick-wall
		low-pass filter, $n = 2048$. FDS: frequency-domain sampling,
		LS-CO: least-squares-optimal constrained out-of-band gain, LDBP: learned
		digital backpropagation, MF: matched filter. }
	\label{fig:results}
\end{figure}

\begin{figure}[b]
	\centering
		\includegraphics{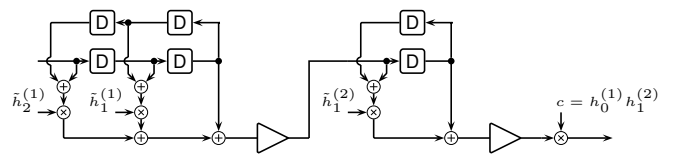}
	\caption{Folded FIR filter implementation for two
	LDBP steps using $h_0^{(i)}$-normalization, i.e.,
	we have $\tilde{h}^{(i)}_j \define h^{(i)}_j/h^{(i)}_0$. The Kerr parameters in
	the nonlinearities have to be scaled accordingly. }
	\label{fig:filter}
\end{figure}

\subsection{Comparison to Few-Step Approaches}


The enhanced SSFM (ESSFM) modifies the nonlinear step based on a
logarithmic perturbation \cite{Secondini2016}. As a result, the
sampled intensity waveform $\{|y_k|^2\}_{k \in \mathbb{Z}}$ is
filtered before applying the nonlinear phase shift. This gives the
same functional form as previous approaches (e.g., \cite{Rafique2011a,
Du2010}), albeit with potentially different performance due to
different choices or heuristics for the filter coefficients used in
the modified nonlinear steps. 

Excluding the overhead due to overlap-and-save techniques, one ESSFM
step requires $4 \log_2 n + 11 + N_c$ real multiplications per sample,
where $2 N_c +1$ is the filter length in the modified nonlinear steps
\cite[Sec.~4, single pol.]{Secondini2016}. We perform 4 ESSFM steps
with $N_c = 20$, which gives roughly the same number of
multiplications as LDBP. The filter coefficients are optimized from
data as suggested in \cite{Secondini2016}. The performance is shown by
the grey line (diamonds) in Fig.~4. The ESSFM achieves a smaller peak
SNR by around $3\,$dB than LDBP. 



\subsection{Deep Learning Interpretation} 

The performance of only the linear steps in LDBP after training
reverts approximately to that of the linear equalizer, as shown by the
dotted green line (crosses) in Fig.~\ref{fig:results}. This leads to
an intuitive interpretation of the task that is accomplished by deep
learning. In particular, the optimized filter coefficients represent
an approximate factorization of the overall linear inverse fiber
response. At first, this may seem trivial because the linear matrix
operator $e^{L \mat{A}}$ can be factored as $e^{\delta \mat{A}}\cdot
\dots \cdot e^{\delta \mat{A}}$ with $L = \delta M$ for arbitrary $M$
to represent shorter propagation distances. However, the factorization
task becomes nontrivial if we also require the individual operators
$e^{\delta \mat{A}}$ to be ``cheap'', i.e., implementable using short
filters.

\begin{remark}
	We also experimented with factoring the $z$-transform polynomial of
	the $75$-tap linear equalizer into a cascade of $3$-tap filters.
	However, this gives no control over the individual filter responses,
	other than the choice of how to distribute the overall gain factor.
	Moreover, it is not obvious how to achieve a good ordering of
	sub-filters in the SSFM. 
\end{remark}

\subsection{Wavelength Division Multiplexing and Higher Baud Rates}


In a WDM system, the performance improvements of ideal single-channel
(or few-channel) DBP are limited due to nonlinear interference from
neighboring channels. This implies that it may be desirable to
sacrifice some accuracy (i.e., target a lower effective SNR), and
further simplify the design of LDBP, e.g., by pruning additional
filter taps. A relaxed accuracy requirement also leaves some margin
for practical impairments such as noise caused by filter coefficient
quantization \cite{Fougstedt2017}. 


The memory introduced by chromatic dispersion increases quadratically
with the considered bandwidth and linearly with the transmission
distance, see \eqref{eq:rule_of_thumb}. For longer links and/or higher
baud rates, this seems to favor frequency-domain equalization (e.g., a
DFT-based linear equalizer) over time-domain equalization in terms of
complexity. On the other hand, the Kerr effect and its compensation
are naturally described in the time domain. One possible approach to
achieve a good performance--complexity trade-off is through digital
sub-band processing. This entails a potential performance loss (due to
possibly uncompensated sub-band interference), but it also reduces the
effective system memory per sub-band. A closer investigation of this
trade-off for LDBP is the subject of ongoing research. 


\fi

\section{Conclusion}
\label{sec:conclusion}

We have considered the problem of reducing the complexity of DBP to
facilitate a real-time DSP implementation. Our approach, called
learned DBP (LDBP), is based on a multi-layer computation graph
generated by the SSFM with many steps. Computational efficiency is
achieved by using, in each step, very short and symmetric FIR filters
that are jointly optimized with deep learning. Numerical results show
that for a single-channel transmission scenario, LDBP can achieve a
favorable performance--complexity trade-off compared to other filter
design methods and perturbation-based ``few-step'' DBP. 

%


\begin{thebibliography}{10}
\providecommand{\url}[1]{#1}
\csname url@samestyle\endcsname
\providecommand{\newblock}{\relax}
\providecommand{\bibinfo}[2]{#2}
\providecommand{\BIBentrySTDinterwordspacing}{\spaceskip=0pt\relax}
\providecommand{\BIBentryALTinterwordstretchfactor}{4}
\providecommand{\BIBentryALTinterwordspacing}{\spaceskip=\fontdimen2\font plus
\BIBentryALTinterwordstretchfactor\fontdimen3\font minus
  \fontdimen4\font\relax}
\providecommand{\BIBforeignlanguage}[2]{{%
\expandafter\ifx\csname l@#1\endcsname\relax
\typeout{** WARNING: IEEEtran.bst: No hyphenation pattern has been}%
\typeout{** loaded for the language `#1'. Using the pattern for}%
\typeout{** the default language instead.}%
\else
\language=\csname l@#1\endcsname
\fi
#2}}
\providecommand{\BIBdecl}{\relax}
\BIBdecl

\bibitem{Agrawal2006}
G.~P. Agrawal, \emph{Nonlinear Fiber Optics}, 4th~ed.\hskip 1em plus 0.5em
  minus 0.4em\relax Academic Press, 2006.

\bibitem{Pare1996}
C.~Par{\'{e}}, A.~Villeneuve, P.-A.~A. B{\'{e}}langer, and N.~J. Doran,
  ``Compensating for dispersion and the nonlinear {Kerr} effect without phase
  conjugation,'' \emph{Optics Letters}, vol.~21, no.~7, pp. 459--461, 1996.

\bibitem{Essiambre2005}
R.-J. Essiambre and P.~J. Winzer, ``Fibre nonlinearities in electronically
  pre-distorted transmission,'' in \emph{Proc. European Conf. Optical
  Communication (ECOC)}, Glasgow, UK, 2005.

\bibitem{Roberts2006}
K.~Roberts, C.~Li, L.~Strawczynski, M.~O'Sullivan, and I.~Hardcastle,
  ``Electronic precompensation of optical nonlinearity,'' \emph{IEEE Photon.
  Technol. Lett.}, vol.~18, no.~2, pp. 403--405, Jan. 2006.

\bibitem{Ip2008}
E.~Ip and J.~M. Kahn, ``Compensation of dispersion and nonlinear impairments
  using digital backpropagation,'' \emph{J. Lightw. Technol.}, vol.~26, pp.
  3416--3425, Oct. 2008.

\bibitem{Du2010}
L.~B. Du and A.~J. Lowery, ``Improved single channel backpropagation for
  intra-channel fiber nonlinearity compensation in long-haul optical
  communication systems.'' \emph{Opt. Express}, vol.~18, no.~16, pp.
  17\,075--17\,088, Jul. 2010.

\bibitem{Shen2011}
T.~S.~R. Shen and A.~P.~T. Lau, ``Fiber nonlinearity compensation using extreme
  learning machine for {DSP}-based coherent communication systems,'' in
  \emph{Proc. Optoelectronics and Communications Conf. (OECC)}, Kaohsiung,
  Taiwan, 2011.

\bibitem{Rafique2011a}
D.~Rafique, M.~Mussolin, M.~Forzati, J.~M{\aa}rtensson, M.~N. Chugtai, and
  A.~D. Ellis, ``Compensation of intra-channel nonlinear fibre impairments
  using simplified digital back-propagation algorithm.'' \emph{Opt. Express},
  vol.~19, no.~10, pp. 9453--9460, Apr. 2011.

\bibitem{Napoli2014}
A.~Napoli, Z.~Maalej, V.~A. J.~M. Sleiffer, M.~Kuschnerov, D.~Rafique,
  E.~Timmers, B.~Spinnler, T.~Rahman, L.~D. Coelho, and N.~Hanik, ``Reduced
  complexity digital back-propagation methods for optical communication
  systems,'' \emph{J. Lightw. Technol.}, vol.~32, no.~7, 2014.

\bibitem{Jarajreh2015}
A.~M. Jarajreh, E.~Giacoumidis, I.~Aldaya, S.~T. Le, A.~Tsokanos,
  Z.~Ghassemlooy, and N.~J. Doran, ``Artificial neural network nonlinear
  equalizer for coherent optical {OFDM},'' \emph{IEEE Photon. Technol. Lett.},
  vol.~27, no.~4, pp. 387--390, Feb. 2015.

\bibitem{Secondini2016}
M.~Secondini, S.~Rommel, G.~Meloni, F.~Fresi, E.~Forestieri, and L.~Poti,
  ``Single-step digital backpropagation for nonlinearity mitigation,''
  \emph{Photon. Netw. Commun.}, vol.~31, no.~3, pp. 493--502, 2016.

\bibitem{Fougstedt2017}
C.~Fougstedt, M.~Mazur, L.~Svensson, H.~Eliasson, M.~Karlsson, and
  P.~Larsson-Edefors, ``Time-domain digital back propagation: Algorithm and
  finite-precision implementation aspects,'' in \emph{Proc. Optical Fiber
  Communication Conf. (OFC)}, Los Angeles, CA, 2017.

\bibitem{Fougstedt2017b}
C.~Fougstedt, L.~Svensson, M.~Mazur, M.~Karlsson, and P.~Larsson-Edefors,
  ``Finite-precision optimization of time-domain digital back propagation by
  inter-symbol interference minimization,'' in \emph{Proc. European Conf.
  Optical Communication}, Gothenburg, Sweden, 2017.

\bibitem{Monterola2001}
C.~Monterola and C.~Saloma, ``Solving the nonlinear {Schroedinger} equation
  with an unsupervised neural network,'' \emph{Opt. Express}, vol.~9, no.~2,
  pp. 72--84, Jul. 2001.

\bibitem{LeCun2015}
Y.~LeCun, Y.~Bengio, and G.~Hinton, ``Deep learning,'' \emph{Nature}, vol. 521,
  no. 7553, pp. 436--444, 2015.

\bibitem{Gregor2010}
K.~Gregor and Y.~Lecun, ``Learning fast approximations of sparse coding,'' in
  \emph{Proc. Int. Conf. Mach. Learning}, 2010.

\bibitem{Lin2017}
H.~W. Lin, M.~Tegmark, and D.~Rolnick, ``Why does deep and cheap learning work
  so well?'' \emph{J. Stat. Phys.}, vol. 168, no.~6, 2017.

\bibitem{Haeger2018ofc}
C.~H{\"{a}}ger and H.~D. Pfister, ``Nonlinear interference mitigation via deep
  neural networks,'' in \emph{Proc. Optical Fiber Communication Conf. (OFC)},
  San Diego, CA, 2018.

\bibitem{Savory2008}
S.~J. Savory, ``Digital filters for coherent optical receivers,'' \emph{Opt.
  Express}, vol.~16, no.~2, pp. 804--817, 2008.

\bibitem{Zhu2009}
L.~Zhu, X.~Li, E.~Mateo, and G.~Li, ``Complementary {FIR} filter pair for
  distributed impairment compensation of {WDM} fiber transmission,'' \emph{IEEE
  Photon. Technol. Lett.}, vol.~21, no.~5, pp. 292--294, Mar. 2009.

\bibitem{Goldfarb2009}
G.~Goldfarb and G.~Li, ``Efficient backward-propagation using wavelet- based
  filtering for fiber backward-propagation,'' \emph{Opt. Express}, vol.~17,
  no.~11, pp. 814--816, May 2009.

\bibitem{Eghbali2014}
A.~Eghbali, H.~Johansson, O.~Gustafsson, and S.~J. Savory, ``Optimal
  least-squares {FIR} digital filters for compensation of chromatic dispersion
  in digital coherent optical receivers,'' \emph{J. Lightw. Technol.}, vol.~32,
  no.~8, pp. 1449--1456, Apr. 2014.

\bibitem{Sheikh2016}
A.~Sheikh, C.~Fougstedt, A.~{Graell i Amat}, P.~Johannisson,
  P.~Larsson-Edefors, and M.~Karlsson, ``Dispersion compensation {FIR} filter
  with improved robustness to coefficient quantization errors,'' \emph{J.
  Lightw. Technol.}, vol.~34, no.~22, pp. 5110--5117, Nov. 2016.

\end{thebibliography}

\end{document}